\documentclass[prb,twocolumn,showpacs,preprintnumbers,amsmath,amssymb]{revtex4}

\usepackage{graphicx}
\usepackage{dcolumn}
\usepackage{bm}

\begin{document}

\preprint{APS/123-QED}

\title{ Non-invasive detection of the evolution of the charge states of a double dot system.}

\author{A. W. Rushforth}
 \email{awr1001@cam.ac.uk}
\author{C. G. Smith}
\author{M. D. Godfrey}
\author{H. E. Beere}
\author{D. A. Ritchie}
\author{M. Pepper}
 \affiliation{Cavendish Laboratory, Madingley Road, Cambridge, CB3 0HE, England
}

\date{\today}

\begin{abstract}
Coupled quantum dots are potential candidates for qubit systems in quantum computing. We use a non-invasive voltage probe to study the evolution of a coupled dot system from a situation where the dots are coupled to the leads to a situation where they are isolated from the leads. Our measurements allow us to identify the movement of electrons between the dots and we can also identify the presence of a charge trap in our system by detecting the movement of electrons between the dots and the charge trap. The data also reveals evidence of electrons moving between the dots via excited states of either the single dots or the double dot molecule.
\end{abstract}

\maketitle

Semiconductor quantum dots have been proposed as one of the most promising candidates to serve as qubits in quantum computation.\cite{RefA} Such systems have the advantage that they can be integrated easily with existing microelectronics technology.\cite{RefB} In such a system the spin state of an electron on a quantum dot could serve as the qubit. Alternatively, the spatial position of an electron on a coupled double dot system could be used. Recently, Pashkin et al \cite{RefC} have demonstrated the entanglement of two charge qubits in a Josephson circuit. Ultimately, it will be necessary to control and read out the state of the qubit. Additionally, it will be necessary to isolate such a system from the surrounding electron reservoirs in order to reduce decoherence due to electron-electron scattering. Recent measurements have demonstrated the ability to measure the movement of electrons between quantum dots using a 1D ballistic channel as a non-invasive voltage probe.\cite{RefI} Here, we use a similar technique to study how a double dot system evolves from a situation where it is connected to the electron reservoirs to a situation where it is isolated from them.  We are able to identify the movement of electrons onto or off of the system and also from one dot to the other. The data may reveal evidence of the electron tunneling between the dots via symmetric and anti-symmetric states of the double dot molecule. Additionally, we can identify the presence of a charge trap in our system, possibly caused by an impurity, and we observe transitions of electrons to or from this charge trap. The importance of identifying unwanted charging events and distinguishing them from desired transitions has been pointed out recently by Buehler et al.\cite{RefJ}

\begin{figure}[t]
\includegraphics{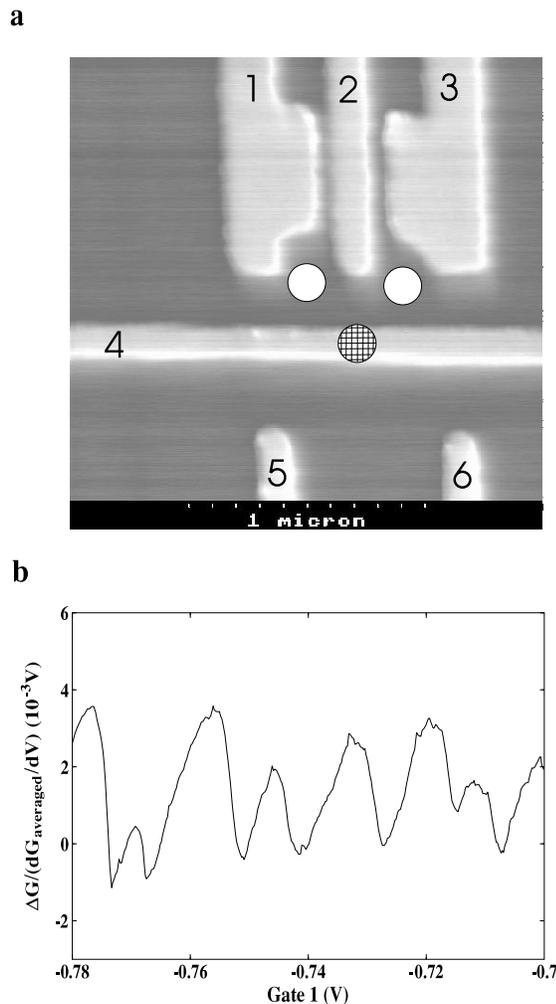}
\caption{\label{fig1:epsart}(a) SEM image of the double dot system, defined by gates 1-4, and the detector, defined by gates 4 and 5. Gate 6 was not used in our measurements. The white circles represent the positions of the quantum dots. The crosshatched circle represents a possible position of a charge trap. (b) The change in the conductance of the detector as a function of the voltage on gate 1.}
\end{figure}

Our device consists of a GaAs/Al$_{x}$Ga$_{1-x}$As heterostructure in which a two-dimensional electron gas (2DEG) layer is formed 90nm below the surface. The gate pattern, fabricated by electron-beam lithography, is shown in Fig. 1(a). The double dot system was formed by the application of negative voltages to gates 1-4. Gate 2 was used to control the coupling between the dots. The detector was defined using gates 4 and 5, with gate 4 also serving to form a tunnel barrier between the double dot system and the detector. We measured the two-terminal conductance of the 1D ballistic channel using standard ac lock-in techniques at 33Hz with a constant excitation voltage of 100$\mu$V. All measurements were made in a dilution refrigerator with a base temperature of 50mK.

The method of non-invasive detection of the potential on a quantum dot was first demonstrated by Field et al.\cite{RefK} Figure 1(b) shows the change in the conductance, $\Delta$G of the detector, which is biased in the tunnelling regime (i.e. $G<2e^{2}/h$), as the voltage on gate 1 is swept. $\Delta$G is the actual detector signal minus a smooth function that fits the detector signal without the steps. This has been divided by the gradient of the smooth function, dG$_{averaged}$/dV to correct for the change in the sensitivity of the detector, as described by Gardelis et al.\cite{RefI}  We observe steps in the conductance, each time an electron leaves the dot system. This is because the width of the ballistic channel is very sensitive to the surrounding electrostatic potential, including that of the dots. The presence of these discrete steps indicates that we are observing Coulomb blockade in the double dot system. Figure 2(a) is a greyscale plot that shows the normalised differential, $(dG/dV)/(dG_{averaged}$$/dV)$, of the detector conductance as a function of the voltages applied to gates 1 and 3, when a double dot is formed. In this plot the dark lines represent the steps in conductance. These features are reproducible between successive sweeps and at different scan speeds ranging from 30 to 360 seconds per linescan. The plot can be separated into four distinct regions labelled A-D. These regions represent the evolution of the system from being coupled to the 2DEG reservoirs to being isolated from them. We shall describe each of these regions in turn.

\begin{figure}[t]
\includegraphics{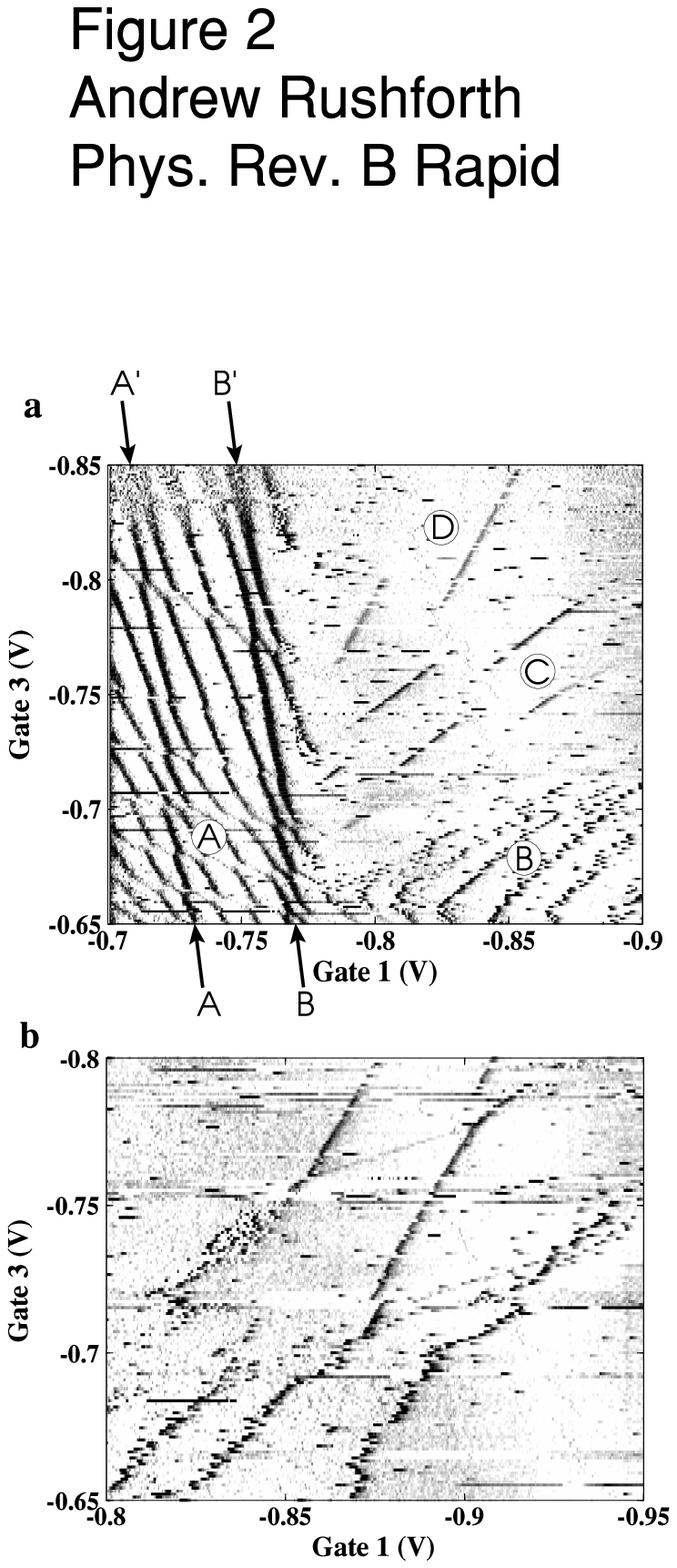}
\caption{\label{fig2:epsart}(a) Greyscale plot showing the normalised differential, $(dG/dV)/(dG_{averaged}$$/dV)$, of the detector conductance as a function of the voltages applied to gates 1 and 3. The black lines represent steps in the detector signal due to electrons entering or leaving the dots. $V_{gate2}$= -0.68V. The arrows (A to A$'$ and B to B$'$) indicate features that are due to a charge trap (see text). (b) As in (a), but $V_{gate2}$= -0.72V.}
\end{figure}

Region A is the regime where the double-dot system is coupled to the 2DEG leads and electrons can move on or off the system via either the left or right tunnel barriers. This is the situation that has been studied by several groups by measuring the conductance through two dots in series.\cite{RefL, RefN} In such measurements conductance peaks are observed at certain points referred to as "triple points". These are represented in Fig. 3 by circles which mark the boundaries between three states of the double dot system that are degenerate in energy. Figure 3 is a schematic diagram showing the stable states of a double dot system in gate voltage space. The numbers in brackets, (n,m) are arbitrarily chosen to represent the electron number on the (left,right) dot respectively. These are not the actual numbers of electrons, which are unknown, but estimated to be in the region of 50 electrons per dot from the geometry and measured charging energy. The black lines represent changes of the total electron number by one. Conductance through the double dot system can only occur at the triple points.\cite{RefN} However, in our system the detector is sensitive to changes in the electron number and so we observe electron transitions all the way along the black lines in Fig. 3 and not just at the triple points. In region A of Fig. 2(a) the near-vertical lines represent an electron leaving or entering the left dot via the left lead. The near-horizontal lines represent an electron leaving or entering the right dot via the right lead. These lines are fainter than the vertical lines because the detector is further away from the right dot and therefore less sensitive to the change in potential on that dot. For the data shown in Fig. 2 the conductance through the double dots is already too small to measure, however this illustrates the advantage of the method of non-invasive detection, in that it allows the system to be studied beyond the regime when the conductance through the dots is immeasurably small. In region A there are also near-vertical features with a steeper gradient. These are highlighted by the arrows (A to A$'$ and B to B$'$) in Fig. 2(a). We believe that these are due to the presence of a charge trap, in the vicinity of the system, that can couple with the double-dot system. We shall return to this issue later in the paper. There are also some features which are perfectly horizontal. These are due to telegraph noise in the detector and are not reproducible.

In region B the voltage on gate 1 is more negative and in this region the barrier between the left dot and the left lead is now too high to allow electrons to traverse the barrier on the time scale of the measurement. It has been demonstrated recently that isolated dots can trap electrons for timescales up to 1000's of seconds.\cite{RefP} In this situation electrons can only escape from the double-dot system via the right lead. We refer to Fig. 3 again to explain why this gives rise to the features in region B. By removing the lines which correspond to electrons moving between the left dot and the left lead, and elongating the lines corresponding to electrons moving between the dots, it can be seen that the charging diagram becomes distorted and consists of a series of lines with a gradient of approximately 1:1 and a period slightly larger than that observed for transitions from the left dot in region A ($\approx$ 15mV). The position of these lines also jumps every time the electron number on the right dot changes by one, giving rise to a zigzag pattern. This is precisely what we observe in region B of Fig. 2(a).

\begin{figure}[floatfix]
\includegraphics{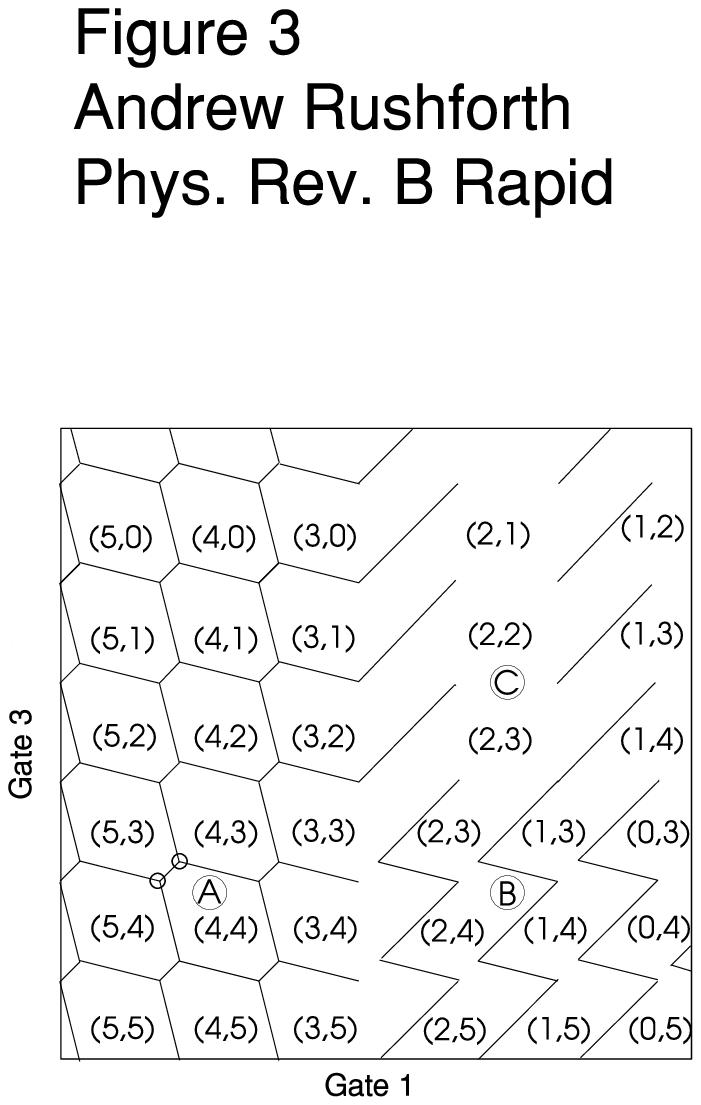}
\caption{\label{fig3:epsart}Schematic representation of the charging diagram showing the different regions with stable charge configuration (n,m) on the (left,right) dot. The circles indicate the position of triple points.}
\end{figure}
In region C both the left and right barriers are too high to allow electrons to tunnel on the timescale of the measurement, and the double dot is completely isolated from the surrounding 2DEG leads. In this situation only transitions between the two dots are possible. Further modification of the charging diagram in Fig. 3 reveals how this affects the position of the observed transitions. We expect to see a series of lines with a gradient close to 1:1, but with a separation larger than that in region B. This appears reasonable when we consider that the capacitance of gate 1 to the left dot alone is 10.7aF and the capacitance of gate 3 to the right dot alone is 8.9aF. Thus $C_{gate 1}$/$C_{gate 3}=1.20$ and the measured gradient in region C is 1.13.

Figure 3 also shows how these lines become discontinuous at certain voltages on Gate 3. This arises because the electron number on the right dot is different when the double dot system first becomes isolated from the leads. This is a consequence of the fact that the measurement is carried out by sweeping the voltage on gate 1 and stepping the voltage on gate 3. Thus, every time the voltage on gate 1 is swept the dot evolves from being open to the leads to being isolated from them and may become isolated with a different electron number, depending upon the voltage on gate 3.

We note that a region, similar to region B, but where only the right dot is isolated from the reservoirs, has also been observed, but is not accessed in the gate voltage range shown in Fig. 2(a). The blurring of features at very negative voltages on gate 3 is not reproducible. This may be due to random switching events, which are believed to be a possible source of apparent dephasing in these systems \cite{RefI}

The pattern observed in region C of Fig. 2(a) is essentially the same as that predicted by Fig. 3. However, region D contains lines with a gradient of 1.98 $\colon$1 indicating that this transition is affected more by gate 1 than gate 3. Figure 2(b) also shows lines with this gradient. This plot reveals that the lines with 1 $\colon$1 gradient split into lines with gradients 1.98 $\colon$1 and 0.35 $\colon$1. We speculate that this is due to the presence of a charge trap, as mentioned earlier. The lines with steep gradient would therefore represent the movement of an electron from the left dot to the charge trap and the lines with shallow gradient represent the electron moving from the charge trap to the right dot. The fact that the line with 1 $\colon$1 gradient appears blurred on approach to this splitting indicates that the probability of a transfer from left to right dot directly is similar to the probability of a transfer via the charge trap. In Fig. 1(a) the crosshatched circle represents a possible position for such a charge trap. We can be confident that this charge trap exists in the 2DEG because conductance measurements through the dots (not shown) and also through the channel formed by gates 1 and 4 alone, reveal Coulomb blockade oscillations with a period corresponding to that identified for the charge trap by the arrows in Fig. 2(a). This is in addition to the Coulomb blockade oscillations corresponding to the dots.

Now, we look in detail at the transitions in region C (Fig. 2(a)). The inset to Fig. 4 shows the conductance step observed when an electron moves from the left to the right dot. It can be seen that a double step is present. This is observed at several positions in region C, but not for the features in region D, where a single step is observed. By differentiating the data we can measure the splitting between the steps. This is shown, in units of energy, in Fig. 4 as a function of perpendicular magnetic field. The voltage axes were calibrated in terms of energy by measuring the charging energy of each dot to be 1meV, using a source-drain bias method.\cite{RefQ} We consider several explanations for the presence of a double step. Firstly, it could be that the electron is transferred between the dots via the charge trap. This is a two stage process that could lead to a double step. However, the double step is observed in regions well away from the split lines observed in Fig. 2(b). The other possibilities are that the electron could be tunnelling into ground and excited states of the second dot, or via symmetric and anti-symmetric states of the double dot molecule. In the former case, the observed behaviour in magnetic field could represent the movement of single particle states, as observed in single dots.\cite{RefR} In the latter case, the general decrease in $\Delta$V could be due to the reduced symmetric anti-symmetric energy gap ($\Delta$SAS) as the electrons become more localised in the dots by the magnetic field. Holleitner et al\cite{RefS} observed the formation of a molecular state when measuring the conductance through two dots in parallel. They measured the magnetic field dependence of the energy gap and observed a structure very similar to that in Fig. 4. They interpreted the dip in $\Delta$V as a transition from a singlet to a triplet state, which arises because the exchange energy is dependent on the overlap of the electron wavefunctions.\cite{RefU} Although symmetric and anti-symmetric states should be formed when the energy levels in the two dots are aligned, it is not clear that the electron should be transferred via both states in a zero temperature model. A possible mechanism for this may be phonon assisted tunneling. Qin et al\cite{RefV} observed an off-resonance conductance in a weakly coupled dot system ($\Delta$SAS=0) which they interpreted as being due to phonon assisted tunneling. In our measurements $\Delta$V lies in the range 30-60$\mu$eV, corresponding to a phonon wavelength\cite{RefW} in the range 310-620nm. This matches the size of our system. Further measurements on similar devices may lead to a clearer explanation of this feature.

\begin{figure}[t]
\includegraphics{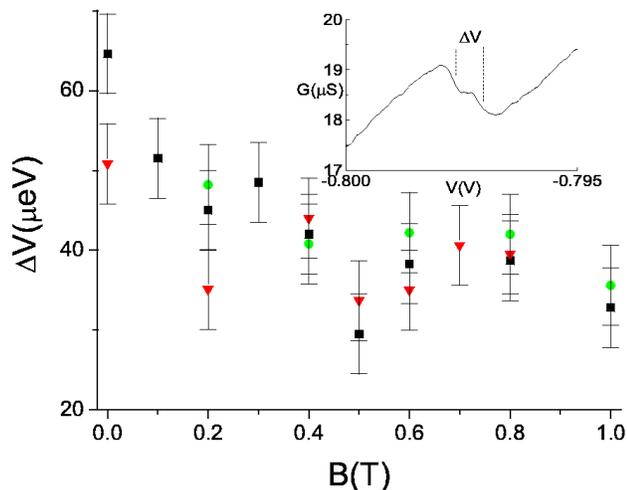}
\caption{\label{fig4:epsart} The separation of the double step as a function of perpendicular magnetic field. Different symbols are for measurements in different parts of region C. Inset$\colon$ An example of the double step feature observed in region C.}
\end{figure}

In conclusion, we have used a non-invasive voltage probe to study the evolution of a double quantum dot system as it evolves from being coupled to the 2DEG leads to being isolated from them. This technique is proving to be increasingly useful in studying isolated systems\cite{RefI, RefP} and can provide information about the presence of charge traps and tunneling via excited states of a dot or molecule. 

\begin{acknowledgments}
We gratefully acknowledge funding by the EPSRC (UK). We also wish to thank Dr Nigel Cooper for useful discussions.
\end{acknowledgments}



\end{document}